\documentstyle[12pt]{article}

\begin{document}

\title{Diagonal deformations of thin center vortices and their stability in Yang-Mills theories}
\author{L. E. Oxman \\ \\ Instituto de F\'{\i}sica, Universidade Federal Fluminense,\\
Campus da Praia Vermelha, Niter\'oi, 24210-340, RJ, Brazil.}
\date{\today}
\maketitle

\begin{abstract}

The importance of center vortices for the understanding of the confining
properties of $SU(N)$ Yang-Mills theories is well established in the lattice.
However, in the continuum, there is a problem concerning the relevance of thick center
vortex backgrounds. They display the so called Savvidy-Nielsen-Olesen
instability, associated with a gyromagnetic ratio $g^{(b)}_m=2$ for the off-diagonal gluons.

In this work, we initially consider the usual definition of a {\it thin} center vortex and rewrite it 
in terms of a local color frame in $SU(N)$ Yang-Mills theories. Then, we define a thick object as a 
diagonal deformation of the thin center vortex. Besides the usual thick background profile, this deformation
also contains a frame defect coupled with gyromagnetic ratio $g^{(d)}_m=1$. 
As a consequence, the analysis of stability is modified. 
In particular, we point out that the defect should stabilize a vortex configuration  
formed by a pair of straight components with opposite fluxes, separated by an appropriate finite distance.

\end{abstract}


{\bf Pacs}: 11.15.-q, 11.15.Tk

\section{Introduction}

The association of magnetic objects with confinement in $SU(N)$ Yang-Mills
theories \cite{3}-\cite{ap} is by now a well established idea. In the last
years, a large amount of information has been accumulated  in the lattice,
identifying center vortices and monopoles as relevant configurations responsible
for confinement \cite{debbio3}-\cite{GKPSZ}. In the lattice, not only a Wilson
loop area law, but also important properties of the confining
string, supported by Monte Carlo simulations and theoretical arguments, has been obtained (for a
review, see ref. \cite{greensite}). 

Various efforts have been devoted to implement in the continuum a formulation
analogous to successful lattice procedures such as center projection and the
maximal center gauge \cite{engelhardt1}-\cite{quandt}, and to derive effective
theories for the monopole sector that can capture large distance physics
\cite{cho-a}-\cite{Shaba}. 

The relevance of magnetic objects in the continuum depends on the stability of
the members in the ensemble when quantum fluctuations are considered around
them. 

In ref. \cite{MS}, the total vacuum energy in a homogeneous color magnetic field has been computed up to one-loop, 
leading to a negative value. This suggested that by the formation of this configuration one could gain energy 
relative to the perturbative ground state.
However, as is well-known, the computed energy also contains an imaginary part, leading to the so
called Savvidy-Nielsen-Olesen instability \cite{sno}-\cite{l-r}.

The argument goes as follows. In the case of $SU(2)$, the one-loop effective
action in the maximally Abelian gauge,
performed around a static magnetic background field  ${\cal B}^3_\mu$ along the
diagonal direction in color space, is given by \cite{cho5},
\begin{eqnarray}
&&\Delta S^{(\rm I)} = \ln \rm{Det}\, K^{(\rm I)}_+  + \ln \rm{Det}\, K^{(\rm
I)}_-
,\nonumber \\
&& K^{(\rm I)}_\pm =-D^2({\cal B}) \pm 2 g \tilde{H}^3_{12}({\cal B}),\nonumber
\\ 
&& D_\mu({\cal B})=\partial_\mu + ig {\cal B}^3_\mu
\makebox[.2in]{,}
\tilde{H}^3_{\mu \nu}({\cal B})=\partial_\mu {\cal B}^3_\nu - 
\partial_\nu {\cal B}^3_\mu .
\label{gm4}
\end{eqnarray}
The eigenvalues for these operators can be thought of as the energy levels for a
particle with charge $g$, spin orientation $\pm$ 1, and gyromagnetic ratio
$g^{(b)}_m=2$. For the operators $K_\pm=-D^2({\cal B}) \pm g_m g B$, when
$B=\tilde{H}^3_{12}({\cal B})$ is homogeneous, these levels are placed at,
\begin{equation}
2gB (n + 1/2) + k^2 \pm g_m gB
\makebox[.2in]{,}
k^2=k_0^2+ k_3^2,
\label{hom}
\end{equation} 
For $n=0$, $B>0$, $K_-$ has eigenvalues at $k^2+(1-g_m) gB$. In the $g_m=2$
case, they are negative for $k^2< gB$, thus originating the
Savvidy-Nielsen-Olesen instability.

Unlike a homogeneous background, the consideration of an ensemble of monopoles
and center vortices would not break Lorentz invariance of the vacuum. 
The total energy  for a thick center vortex background field, including the classical part and considering quantum fluctuations up to one-loop,  
has been computed in refs. \cite{bordag,diakonov-center}. Under quite general conditions, it has been shown that the total energy is always negative. 
However, similarly to what happens in the case of the homogeneous background, the energy contains an imaginary part implying instability. This is
originated due to the presence of bound states for the fluctuation operator. In fact, bound states, with the corresponding negative eigenvalues, are a
characteristic of Schr\"odinger operators for spin one particles with
gyromagnetic ratio $g_m >1$, see refs.  \cite{bordagQM, morozQM}. Then, the
$g^{(b)}_m=2$ gyromagnetic ratio in eq. (\ref{gm4}) is at the root of the
instability problems when center vortices are introduced as background fields.
It is important to underline that in refs. \cite{bordagQM, morozQM} the
discussion is oriented to spin one-half particles, so that the mathematics there
for gyromagnetic ratio two or four, here corresponds to $g_m=1$ or $g_m=2$,
respectively. 

Besides the thick center vortex background field, another concept has been introduced 
in the Literature: the so called {\it thin} center vortex \cite{engelhardt1,reinhardt}. As we will see,
the latter does not correspond to taking the zero radius limit of the former.
In this article, we will initially consider the thin center vortex and rewrite it 
in terms of a local color frame, extending the discussion for
$SU(2)$ given in ref. \cite{lucho}. Next, we will consider a diagonal deformation of the thin center vortex
to define a new type of thick object. This will lead to a different situation regarding
stability, suggesting that our thick center vortices could be relevant degrees of freedom in continuum Yang-Mills theories.

In section \S \ref{cvym}, we discuss the local color frame representation of $SU(N)$ thin center 
vortices and introduce their diagonal deformation. Section \S \ref{oneloop} is devoted to review the 
calculation of the one-loop effective action, emphasising the modifications implied by the new type
of thick center vortex object. In section \S \ref{stability}, we present a careful analysis of stability for the different 
alternatives, while in \S \ref{DC} we discuss the appropriate boundary conditions to be imposed on the charged fields
at the defect locations. Finally, in section \S \ref{conc}, we present our conclusions.

\vspace{.2in}

\section{Center vortices in Yang-Mills theories}
\label{cvym}

As usual, the Yang-Mills action is given by 
\[
S_{YM}=\frac{1}{4}\int d^4 x\; \vec{F}_{\mu \nu}\cdot \vec{F}_{\mu \nu},
\]
\begin{equation}
\vec{F}_{\mu \nu} \cdot \vec{T}=(i/g)\left[D_\mu,D_\nu \right]
\makebox[.3in]{,}
D_\mu=\partial_\mu-ig \vec{A}_{\mu}\cdot \vec{T},
\label{field-strength}
\end{equation}
where $T^A$, $A=1,..,N^2-1$ are hermitian generators of $SU(N)$ satisfying,
$\left[ T^{A},T^{B}\right]=if^{ABC}T^C$, $tr\, (T^A
T^B)=\frac{1}{2}\delta^{AB}$.
For $SU(2)$, they can be realized as $T^A=\tau^A/2$, where $\tau^A$ are the
Pauli matrices, and the structure constants $f^{ABC}$ are given by the
Levi-Civita symbol $\epsilon^{ABC}$. For $SU(3)$, $T^A=\lambda^A/2$, where
$\lambda^A$
are the Gell-mann matrices. We will denote the color indices along diagonal and
off-diagonal directions as $i$ and $a$, respectively. 

\subsection{Thick center vortices: usual procedure}

A thick center vortex along the Cartan subalgebra of $SU(N)$ is usually
introduced as a background field,
\begin{equation}
\vec{A}^{{\,{\rm I}}}_\mu =({\cal B}^i_\mu + q^i_\mu)\, \hat{e}_i + q^a_\mu\,
\hat{e}_a,
\label{thickv}
\end{equation}
where $\hat{e}_A$ is the spacetime independent canonical basis in color space,
and $q_\mu^A$ represents perturbative quantum fluctuations.

The generator,
\begin{equation}
 T^{r}=\sqrt{\frac{N}{2(N-1)}}E 
\makebox[.5in]{,} r=N^2-1, 
\end{equation}
with $E$ a diagonal matrix having components,
\begin{equation}
E_{\alpha \alpha} =\left\{ \begin{array}{ll}
(1/N),  & \alpha =1, \dots, N-1 \\
(1/N)-1,& \alpha =N,
\end{array}\right. 
\label{diagE}
\end{equation}
is in the Cartan subalgebra. Then, a thick center vortex, placed at the $x_3$-axis
for every Euclidean time $x_0$, can be introduced by eq. (\ref{thickv}), together with
the profile
\begin{equation}
{\cal B}^i_\mu = \frac{1}{g}\sqrt{2(N-1)/N} f(\rho) \partial_\mu \varphi\,
\delta^{ir},
\label{thickv'}
\end{equation}
where $\rho$, $\varphi$ are polar coordinates in the $x_1$, $x_2$ plane. The
profile function satisfies, $f(\rho)\to 0$, for $\rho \to 0$, while $f(\rho)=1$,
for $\rho > \rho_v$. This manner to introduce a center vortex will be denoted
as procedure I, the associated instability problem has been studied in refs.
\cite{nn-no,l-r,cho5}, \cite{bordag,diakonov-center}.

\subsection{Thin center vortices: usual procedure}

Now, in the continuum, closed thin center vortices have been
introduced on top of a field ${\cal A}^A_\mu\, T^A$, by proposing the
configuration,
\begin{equation}
\vec{A}^{\,\rm{thin}}_\mu \cdot \vec{T} = S {\cal A}^A_\mu\, T^A
S^{-1}+\frac{i}{g}
S\partial_\mu S^{-1}-I_\mu (\vartheta),
\label{usual-thin}
\end{equation} 
$S\in SU(N)$ \cite{engelhardt1,reinhardt}.
When crossing a three-volume $\vartheta$ (ideal vortex) whose border gives the
closed
thin center vortex worldsheet, the mapping $S$ changes to $e^{\pm i2\pi/N}\, S$.
The term $I_\mu(\vartheta)$ is concentrated on $\vartheta$ and is designed to compensate
derivatives of the discontinuity of $S$ at $\vartheta$, thus subtructing the
ideal part, only keeping the thin center vortex part. In the next subsection we will show how this procedure works
for a simple example. 

It can be seen that the Wilson loop,
\begin{equation}
 W({\cal C})=\frac{1}{2}\, P \left\{ e^{ig \oint dx_\mu\, \vec{A}_\mu \cdot \vec{T}}\right\},
\end{equation}
where $P\{ \cdot \}$ stands for path ordering, when computed for $\vec{A}^{\,\rm{thin}}_\mu$ differs from the one for $\vec{{\cal
A}}_\mu$ by a factor $e^{i2\pi/N}$, when the Wilson loop encircles the closed thin center vortex once (see ref. \cite{{engelhardt1}}).

In order to put the thin field configurations in eq. (\ref{usual-thin}) in a simpler form, permitting its
extension to thick objects, we will introduce a local basis $\hat{n}_A$ in color space, 
\begin{equation}
S T^A S^{-1}= \hat{n}_A \cdot \vec{T}
\makebox[.5in]{,} 
\hat{n}_A=R_{\hat{n}}\, \hat{e}_A ,
\label{frame}
\end{equation}
and define,
\begin{equation}
C^{(\hat{n})A}_\mu = -\frac{1}{gN} f^{ABC} \hat{n}_B \cdot \partial_\mu
\hat{n}_C .
\label{CAm}
\end{equation}
The adjoint representation $R=R_{\hat{n}}$ is generated by the
matrices $M^A$, with elements $(M^A)^{BC}=-i f^{ABC}$, that satisfy $\left[
M^{A},M^{B}\right]=if^{ABC}M^C$, $tr\, (M^A M^B)= N \delta^{AB}$. 
On the other hand, in eq. (\ref{CAm}) we have, 
\begin{equation}
\hat{n}_B \cdot \partial_\mu
\hat{n}_C=\hat{e}_B \cdot R^{-1}\partial_\mu R\, \hat{e}_C=(R^{-1}\partial_\mu R)_{BC},
\end{equation}
where we can write,
\begin{equation}
 R^{-1}\partial_\mu R =iX^A_\mu \, M^A,
\label{Rg}
\end{equation}
as the first member is in the algebra. Then, we obtain,
\begin{equation}
 \hat{n}_B \cdot \partial_\mu \hat{n}_C=f^{ABC} X^A_\mu,
\label{enedotene}
\end{equation}
which together with
eq. (\ref{CAm}), and the property $f^{ACD} f^{BCD}= N \delta^{AB}$, implies
$X^A_\mu = -g C^A_\mu$. Thus, replacing in eqs. (\ref{Rg}) and (\ref{enedotene}), we get,
\begin{equation}
C^{(\hat{n})A}_\mu \, M^A = \frac{i}{g} R_{\hat{n}}^{-1}\partial_\mu
R_{\hat{n}}
\makebox[.5in]{,}
\hat{n}_B \cdot \partial_\mu \hat{n}_C = -gf^{ABC} C^{(\hat{n})A}_\mu.
\label{equali}
\end{equation}
From the first equality, it seems we can also write $C^A_\mu T^A=\frac{i}{g}
S^{-1}\partial_\mu S$, or $-C^A_\mu\, \hat{n}_A \cdot
\vec{T}=\frac{i}{g} S \partial_\mu S^{-1}$. However, unlike $S$, the adjoint
representation $R$ and the associated frame $\hat{n}_A$ are always
continuous, so that $C^A_\mu\, \hat{n}_A$ contains no term concentrated on
$\vartheta$. Then, the correct expression is,
\begin{equation}
 -C^A_\mu\, \hat{n}_A \cdot \vec{T}=\frac{i}{g} S \partial_\mu S^{-1}
-I_\mu(\vartheta), 
\end{equation}
that when replaced in eq. (\ref{usual-thin}), together with $S {\cal A}^A_\mu\, T^A S^{-1}={\cal A}^A_\mu\, \hat{n}_A \cdot \vec{T}$, 
leads to an equivalent representation of the thin configuration proposed in refs. \cite{engelhardt1,reinhardt},
\begin{equation}
\vec{A}^{\,\rm{thin}}_\mu =(-C^{(\hat{n})A}_\mu +{\cal A}^A_\mu)\, \hat{n}_A  ,
\label{ansatz}
\end{equation}
where the thin center vortices are encoded as defects of a local color frame
(for $SU(2)$, see ref. \cite{lucho}). Here, the field $\vec{{\cal A}}={\cal A}^A_\mu\,
\hat{e}_A$ represents a trivial perturbative sector that includes quantum
fluctuations. 

\subsection{Diagonal deformation of a thin center vortex}
\label{diagdef}

Now, we will see that the usual manner to introduce a thin center
vortex (eq. (\ref{usual-thin}) or (\ref{ansatz})) does not correspond to considering the limit
$\rho_v \to 0$ in the usual definition for a thick center vortex in eq.
(\ref{thickv}). Taking this limit in (\ref{thickv}), we would obtain,
\begin{equation}
 \vec{A}^{{\,{\rm I}}}_\mu \to \left[ \frac{1}{g}\sqrt{2(N-1)/N}\, \partial_\mu \varphi \,
\delta^{ir}+ q^i_\mu \right]\, \hat{e}_i + q^a_\mu\, \hat{e}_a.
\label{tlimit}
\end{equation}
On the other hand, consider the usual procedure to introduce a thin center vortex, given in eq. (\ref{usual-thin}), for 
the example where the mapping is defined by, 
\begin{equation}
S=e^{i\varphi \, E}=e^{i\varphi \sqrt{2(N-1)/N}\, T^r}.
\end{equation}
In this case, we have,
\begin{equation}
 \frac{i}{g} S \partial_\mu S^{-1} = \frac{1}{g} \, \partial_\mu \varphi \, E +\frac{2\pi}{g}
\theta (x_1) \delta (x_2)\, E \, \delta_{\mu 2},
\label{Sder}
\end{equation}
where $\partial_\mu \varphi$ is defined as usual,
\begin{equation}
 \partial_1 \varphi = -\frac{x_2}{x_1^2+x_2^2}
\makebox[.5in]{,}
 \partial_2 \varphi = \frac{x_1}{x_1^2+x_2^2},
\end{equation}
and the second term in eq. (\ref{Sder}) is originated when deriving the discontinuity of $S^{-1}$, that is concentrated on the spacetime points 
with $x_2=0$, positive $x_1$, and any $x_0,x_3$, which define a three-volume $\vartheta$. 
In this regard, note that when approaching $\vartheta$ from the $x_2 >0$ ($x_2 <0$) side, we have $\varphi \to 0$ ($\varphi \to 2\pi$)
and $S(0)\to I$ ($S(2\pi)\to e^{i2\pi/N}\, I$). Then, $I_\mu(\vartheta)$ in eq. (\ref{usual-thin}) must be defined as,
\begin{equation}
 I_\mu(\vartheta)= \frac{2\pi}{g} \theta (x_1) \delta (x_2)\, E\, \delta_{\mu 2},
\end{equation}
and using the first equation in (\ref{frame}), we have,
\begin{equation}
\vec{A}^{\,\rm{thin}}_\mu = \left[ \frac{1}{g} \sqrt{2(N-1)/N}\, \partial_\mu \varphi\,
\delta^{ir} + {\cal A}^i_\mu\right] \, \hat{e}_i + {\cal
A}^a_\mu\,  
\hat{n}_a.
\label{examp-simp}
\end{equation}
Equivalently, we note that in this example the local frame can be obtained by applying (cf. eq. (\ref{frame}))
\begin{equation}
 R=e^{i\varphi \sqrt{2(N-1)/N}\, M^r} ,
\end{equation}
on the basis $\hat{e}_A$. As the mapping $R$ has no discontinuity on $\vartheta$, as it is in the adjoint representation, 
when computing the first equality in (\ref{equali}) no term concentrated on $\vartheta$ arises, and
we simply get,
\begin{equation}
C^{(\hat{n}) i}_\mu = -\frac{1}{g} \sqrt{2(N-1)/N}\, \partial_\mu \varphi\,
\delta^{ir}
\makebox[.2in]{,}
C^{(\hat{n}) a}_\mu =0.
\label{examp}
\end{equation}
That is, when working with the representation (\ref{ansatz}), the ideal vortex is naturally eliminated.  
Finally, replacing (\ref{examp}) in (\ref{ansatz}), we reobtain (\ref{examp-simp}).

The fields $q^A_\mu$ in eq. (\ref{tlimit}) as well as the fields ${\cal A}^A_\mu$ in eq. (\ref{examp-simp}) represent trivial quantum fluctuations, 
so that there is an essential difference between these parametrizations. The off-diagonal sectors are given by $q^a_\mu\, \hat{e}_a$ and ${\cal A}^a_\mu\, \hat{n}_a$, respectively, so that, while in the former case trivial fluctuations are accompanied by global frame elements $\hat{e}_a$, in the latter, they are accompanied by local elements $\hat{n}_a$ containing defects. Indeed, note that for $N=2$, we have
$R=e^{i\varphi M^3}$, that is, when we go around the
vortex, the elements $\hat{n}_1$, $\hat{n}_2$ rotate once, leaving $\hat{n}_3$ invariant. For
$N=3$, we have $R=e^{i\varphi \frac{2}{\sqrt{3}}\, M^8}$, and using that the
nontrivial structure constants involving the eigth color are,
$f^{458}=f^{678}=\sqrt{3}/2$, we see that in this case 
the pair of basis elements $\hat{n}_4$, $\hat{n}_5$ and $\hat{n}_6$, $\hat{n}_7$
rotate once, leaving $\hat{n}_A$, $A=1,2,3,8$ invariant.
  
The previous discussion suggests that a new type of thick object can be naturally introduced, on top of a field
${\cal A}^A_\mu\, \hat{e}_A$, by including a profile function $f(\rho)$ in the first term of the parametrization (\ref{examp-simp}).
This object will be called a diagonal deformation of the thin center vortex, and can be written in terms of the profile ${\cal B}^i_\mu$ defined in eq. (\ref{thickv'}), 
\begin{equation}
\vec{A}^{\,{\rm II}}_\mu =({\cal B}^i_\mu + {\cal A}^i_\mu)\, \hat{e}_i + {\cal
A}^a_\mu\,
\hat{n}_a
\makebox[.3in]{,}
\hat{n}_a= e^{i\varphi
\sqrt{\frac{2(N-1)}{N}}\, M^r}\, \hat{e}_a .
\label{thickvII}
\end{equation} 
The frame defect in this new object is expected to be nontrivialy manifested. In this regard, we would like 
to emphazise again that $q^a_\mu$ in eq. (\ref{thickv}), as well as ${\cal A}^a_\mu$ in eq. (\ref{thickvII}), represent perturbative 
sectors so that no defects are to be attributed to these components. In other words, the ans\"atze in eqs. (\ref{thickv}) and
(\ref{thickvII}) describe inequivalent objects. 
Note also that in
the limits $q^A_\mu \to 0$, ${\cal A}^A_\mu \to 0$, both expressions
(\ref{thickv}) and 
(\ref{thickvII}) coincide. Then, around these points, for a loop passing outside
the center vortex cores, in both procedures the Wilson loop $W$ is close to a
center element. An essential difference is that 
in the second procedure, because of the local color directions $\hat{n}_a$, $W$
enjoys an interesting property. For any loop linking the diagonally deformed center vortex
worldsheet (passing up to a distance $\rho_v$), 
$W[\vec{A}^{\,{\rm II}}]$ and $W[\vec{{\cal A}}]$ always differ by a center
element, as in that region $\vec{A}^{\,{\rm II}}$ coincides with the thin center vortex in eq. (\ref{usual-thin}).
In the following sections, differences between the usual 
thick center vortex background field, and the diagonal deformation of a thin center vortex will be further analyzed.

\section{One-loop effective action}
\label{oneloop}

To compute and compare the field strength tensors for $\vec{A}^{{\,{\rm
I}}}_\mu$ and $\vec{A}^{\,{\rm II}}_\mu$ in eqs. (\ref{thickv}) 
and (\ref{thickvII}), we will consider a general field of the form,
\begin{equation}
\vec{A}_\mu = Y^A_\mu\, \hat{u}_A
\makebox[.5in]{,}
\hat{u}_A=R_{\hat{u}}\, \hat{e}_A ,
\end{equation}
and then will make the appropriate replacements.
That is,
\begin{equation}
Y^i_\mu ={\cal B}^i_ \mu +Q^i_\mu\ \makebox[.5in]{,} 
Y^a_\mu = Q^a_\mu,
\end{equation}
where in order to obtain cases {\rm I} an {\rm II}, we will have to consider,
\begin{equation}
\begin{array}{ll}
{\rm I})~~ Q^A_\mu= q^A_\mu,\; & \hat{u}_A=\hat{e}_A,   \\
{\rm II})~ Q^A_\mu= {\cal A}^A_\mu,\; & \hat{u}_i=\hat{e}_i,~ \hat{u}_a =
\hat{n}_a.
\end{array}
\label{IandII}
\end{equation}
From eq. (\ref{field-strength}) and $[\hat{u}_A\cdot \vec{T}, \hat{u}_B \cdot
\vec{T}]=i f^{ABC} \hat{u}_C \cdot \vec{T}$, we get, 
\begin{eqnarray}
\vec{F}_{\mu \nu}=[\partial_\mu Y^A_\nu-\partial_\nu Y^A_\mu]\, \hat{u}_A  +
Y^A_\nu(\hat{u}_C \cdot \partial_\mu \hat{u}_A) \hat{u}_C &&\nonumber \\
-Y^A_\mu (\hat{u}_C \cdot \partial_\nu\hat{u}_A) \hat{u}_C +g f^{ABC} Y^A_\mu
Y^B_\nu \hat{u}_C .&&
\label{inter}
\end{eqnarray}
Using the second equality in eq. (\ref{equali}), this can be rewritten in terms
of $C^A_\mu=C^{(\hat{u})A}_\mu$, given by eq. (\ref{CAm}) with $\hat{u}_A$
in the place of $\hat{n}_A$. In case II, this will give a nontrivial contribution of the 
local color frame in eq. (\ref{thickvII}). 
 
In this manner, we obtain,
\begin{equation}
\vec{F}_{\mu \nu}=G^A_{\mu \nu}\, \hat{u}_A
\makebox[.5in]{,}
S_{YM}=\int d^4x\, \frac{1}{4} G^A_{\mu \nu}G^A_{\mu \nu},
\end{equation}
\begin{eqnarray}
G^A_{\mu \nu} &=& {\cal F}^A_{\mu \nu}(Y)+g f^{ABC}(C^{B}_\mu Y^C_\nu-C^{B}_\nu
Y^C_\mu) \nonumber \\
&=& {\cal F}^A_{\mu \nu}(Y+C)-{\cal F}^A_{\mu \nu}(C),
\label{geA}
\end{eqnarray}
where ${\cal F}^{A}_{\mu \nu }(Y)=\partial_{\mu }Y_{\nu}^{A}-\partial _{\nu}Y_{\mu}^{A}
+ g f^{ABC}Y_{\mu }^{B} Y_{\nu }^{C}$. That is, as for our examples in eq. (\ref{IandII}) the off-diagonal fields $C^a_\mu$ vanish (cf. eq. (\ref{examp})), we obtain,
\begin{equation}
G_ {\mu \nu}^i = \partial_\mu Y_\nu^i - \partial_\nu Y_\mu^i + g f^{ibc} Q_\mu^b Q_\nu^c ,
\label{Gmunu}
\end{equation}
\begin{eqnarray}
\lefteqn{{\cal F}^a_{\mu \nu}(Y+C) = D_\mu^{ab} Q^b_\nu - 
D_\nu^{ab} Q^b_\mu + g
f^{abc} Q^b_\mu Q^c_\nu ,
\label{off-diag}} \nonumber \\
&& D_\mu^{ab}= D_\mu^{ab}(Y+C)=\delta^{ab}\partial_{\mu}- g f^{abi}
(Y^{i}_\mu+C^{i}_\mu).\nonumber \\
\label{Defi}
\end{eqnarray}
In general, we can write, 
\begin{equation}
{\cal F}^A_{\mu \nu}(C)\, M^A =(i/g)\left[{\cal D}_\mu, {\cal D}_\nu \right]
\makebox[.5in]{,} 
{\cal D}_\mu=\partial_\mu-ig C^{A}_\mu M^A,
\end{equation}
so that eq.
(\ref{equali}) implies,
\begin{equation}
{\cal F}^A_{\mu \nu}(C) = (i/gN)\, tr\, (M^A
R_{\hat{u}}^{-1}[\partial_\mu,\partial_\nu]R_{\hat{u}}),
\label{FC}
\end{equation}
that is nonzero only in places where $R_{\hat{u}}$ contains defects. Of course,
as in case I the frame is global, 
${\cal F}^A_{\mu \nu}(C)$ vanish. 
In case II, for our example, as $R_{\hat{n}}$ is a rotation along the diagonal
direction
$M^r$, we have ${\cal F}^a_{\mu \nu}(C)=0$.  This property is also satisfied by
other frames describing 
monopoles and correlated center vortices, when monopoles are encoded as defects
of $\hat{n}_i$. For instance, in the case of $SU(2)$, this happens
whenever monopoles are described as defects of $\hat{n}_3$ \cite{lucho}. 
In all these cases we have $G^a_{\mu \nu}={\cal F}^a_{\mu \nu}(Y+C)$ and,
considering $Q^A_\mu$ as quantum 
fluctuations, a standard calculation shows that up to quadratic terms, we have,
\begin{eqnarray}
\frac{1}{4} G^{a}_{\mu \nu}G^{a}_{\mu \nu }|_{f} &=& -\frac{1}{2} Q_{\nu }^{c}
D^{ca}_\mu D^{ab}_\mu Q_{\nu }^{b} +\frac{1}{2} g f^{ibc} \tilde{H}^i_{\mu \nu}
Q_{\mu }^{b}Q_{\nu }^{c},
\label{Gmunua} \nonumber \\
\tilde{H}^i_{\mu \nu}&=&\tilde{H}^i_{\mu \nu}({\cal B}) +\tilde{H}^i_{\mu
\nu}(C),
\label{Htilde}
\end{eqnarray}
where
\begin{equation}
\tilde{H}^i_{\mu \nu}({\cal B}) =\partial_{\mu}{\cal B}^i_\nu -\partial_{\nu}{\cal B}^i_\nu
\makebox[.5in]{,}
D_\mu^{ab}= D_\mu^{ab}({\cal B}+C).
\end{equation} 
Here we have used the maximally Abelian gauge condition, $D^{ab}_\mu Q_{\mu
}^{b} = 0$, and the property of the structure constants $f^{abi} f^{bcj} +
f^{abj} f^{bic} =0$. In addition, up to quadratic terms involving off-diagonal
fluctuations originated from $G^i_{\mu \nu}$ we have
\begin{eqnarray}
\frac{1}{4}G^i_{\mu \nu}G^i_{\mu \nu} |_{f} &=& \frac{1}{2} g f^{ibc}
\tilde{H}^i_{\mu \nu}({\cal B})\, Q_{\mu}^{b} Q_{\nu}^{c}.
\label{fluct}
\end{eqnarray}
Collecting these terms, we get,
\begin{eqnarray}
\frac{1}{4} G^{A}_{\mu \nu}G^{A}_{\mu \nu}|_{f}=-(1/2)\, Q_{\nu }^{c}
D^{ca}_\mu({\cal B}+C) D^{ab}_\mu({\cal B}+C) Q_{\nu }^{b}&&
\nonumber \\
+(1/2)\, g f^{ibc} [2 \tilde{H}^i_{\mu \nu}({\cal B})+\tilde{H}^i_{\mu \nu}(C)] Q_{\mu }^{b}Q_{\nu }^{c}.&& \nonumber \\
\label{summa}
\end{eqnarray}
For $SU(2)$, the contribution of the terms in eq. (\ref{summa})
to the
Yang-Mills action in case I or II can be written in the form,
\begin{eqnarray}
&& S^f_{YM} = \int d^4x\, [- \bar{\Phi}_\mu D^2({\cal B}+C) \Phi_\mu - i
\Upsilon_{\mu \nu} \bar{\Phi}_\mu \Phi_\nu ], \nonumber \\
&& \Phi_\mu = \frac{1}{\sqrt{2}} (Q_{\mu }^{1}+i Q_{\mu }^{2})
\makebox[.2in]{,}
D_\mu({\cal B}+C) =\partial_\mu + i g ({\cal B}^3_\mu+C^{3}_\mu),\nonumber \\
&&\Upsilon_{\mu \nu}= g\,[2 \tilde{H}^3_{\mu \nu}({\cal B})+\tilde{H}^3_{\mu
\nu}(C)], 
\label{brec}
\end{eqnarray}
and the MAG gauge fixing condition is $D_\mu \Phi_\mu=0$, $\bar{D}_\mu
\bar{\Phi}_\mu=0$. 

The one-loop calculation of the path-integral over
off-diagonal fluctuations and ghosts is \cite{trick},
\begin{equation}
\exp (-\Delta S) = {\rm{Det}}^{-1} (- D^2\, {\mathbf I} -
i{\mathbf\Upsilon})\, {\rm{Det}}^2 (-D^2),
\end{equation}
where the second
factor comes from the Fadeev-Popov determinant. This implies,
\begin{equation}
\Delta S = 2\ln \rm{Det}\, (- D^2) +\ln \rm{Det}\, ({\mathbf I} +i D^{-2}
{\mathbf \Upsilon}),
\label{deltab}
\end{equation}
where one uses, $\ln \rm{Det}\, (- D^2\, {\mathbf I})= 4 \ln \rm{Det}\, (-
D^2)$. 
The index structure has been worked out in ref. \cite{trick}, using 
\[
\ln \rm{Det}\, ({\mathbf I} +i D^{-2} {\mathbf \Upsilon}) = -\sum_{n=1} (1/n)
\rm{Tr}\, [-i D^{-2} {\mathbf \Upsilon}]^n.
\]
For a static background along $\hat{z}$, the only nonzero components of
${\mathbf \Upsilon}$ are $\Upsilon_{12}=-\Upsilon_{21}=\Upsilon$. For odd $n$ the
trace above is zero,
while for even $n$,
\begin{equation}
\rm{Tr}\, [-i D^{-2} {\mathbf \Upsilon}]^n= 2(-1)^{\frac{n}{2}} \rm{Tr}\, [i
D^{-2}\Upsilon]^n, 
\end{equation}
so that one obtains, 
\[
\ln \rm{Det} ({\mathbf I} + i D^{-2} {\mathbf \Upsilon})=\ln \rm{Det}
(I-D^{-2}\Upsilon)+ \ln \rm{Det} (I+D^{-2}\Upsilon),
\]
\begin{equation}
\Delta S = \ln \rm{Det}\, K_+  + \ln \rm{Det}\, K_-
\makebox[.15in]{,}
K_\pm=(-D^2 \pm \Upsilon).
\label{effectiveS}
\end{equation}
For procedure I, as the frame is
global we have $C^{(\hat{u})A}_\mu=0$. Then, replacing eq. (\ref{brec}) in (\ref{effectiveS}), the
effective action
$\Delta S^{(\rm{I})}$ displayed in eq. (\ref{gm4}) is obtained, containing the
Schr\"odinger operators,
\begin{equation}
K^{(\rm{I})}_\pm 
= - D^2({\cal B}) \pm \Upsilon^{(\rm{I})}
\makebox[.3in]{,}  
\Upsilon^{(\rm{I})}=2g\, \tilde{H}^3_{12}({\cal B})
\makebox[.5in]{,}
{\cal B}^3_\mu=\frac{1}{g} f(\rho)\, \partial_\mu \varphi .
\label{BKI}
\end{equation}
In procedure II, using again eq. (\ref{brec}), we arrive at an effective action
$\Delta S^{(\rm{II})}$, obtained by considering in eq.
(\ref{effectiveS}) the operators,
\begin{equation}
K^{(\rm{II})}_\pm = - D^2({\cal B}+C^{(\hat{n})}) \pm \Upsilon^{(\rm{II})}
\makebox[.3in]{,}
\Upsilon^{(\rm{II})}=g\,
[2\tilde{H}^3_{12}({\cal
B})+\tilde{H}^3_{12}(C^{(\hat{n})})],
\label{SchC}
\end{equation}
with ${\cal B}^3_\mu$ given as in eq. (\ref{BKI}) and
$C_\mu^{(\hat{n})3}=-\frac{1}{g}\partial_\mu \varphi$, thus implying,
\begin{equation} 
\tilde{H}^3_{12}(C^{(\hat{n})})=-\frac{1}{g}[\partial_1,\partial_2]\varphi =
-\frac{2\pi}{g} \delta^{(2)}(x,y).
\end{equation}
For $SU(3)$, the fields ${\cal B}^i_\mu$, $C^{(\hat{n})i}_\mu$ in eqs.
(\ref{thickv'}),
(\ref{examp}) are along the eigth-direction. As the structure constants
$f^{8bc}$ are only nonzero for $bc$ taking values $\{ 4, 5\}$ or $\{ 6, 7\}$,
we see from eq. (\ref{Defi}) that the only fluctuations that are coupled with
them in eq. (\ref{summa}) can be grouped in two terms containing either the indices 
$\{ 4, 5\}$ or $\{ 6, 7\}$. Using $f_{458}=f_{678}=\frac{\sqrt{3}}{2}$,
and
\begin{equation}
\frac{\sqrt{3}}{2}{\cal B}^8_\mu =f(\rho)\,
\frac{1}{g}\partial_\mu \varphi
\makebox[.5in]{,}
\frac{\sqrt{3}}{2}C^{(\hat{n})8}_\mu=-\frac{1}{g}\partial_\mu
\varphi ,
\end{equation} 
(cf. eqs. (\ref{thickv'}), (\ref{examp})), we see that each sector, 
\begin{equation}
\Phi^{1}_\mu = \frac{1}{\sqrt{2}} (Q_{\mu }^{4}+iQ_{\mu }^{5})
\makebox[.5in]{,} \Phi^{2}_\mu
= \frac{1}{\sqrt{2}} (Q_{\mu }^{6}+iQ_{\mu }^{7}),
\end{equation} 
gives a contribution to the effective action like in $SU(2)$ (see also ref. \cite{effqcd}). 
Then, it is clear that stability properties are common to $SU(2)$ and $SU(3)$.

\section{Stability Analysis}
\label{stability}

In what follows we will be concerned with the existence of bound states in the
case of (Euclidean) time-independent configurations with translation symmetry
along the $x_3$-direction. Then, we can separate the ($x_0$, $x_3$)
from the $(x_1, x_2)$ variables. To simplify the
discussion we will only write the equations and wave functions for the separated
$(x_1,x_2)$ sector. The associated eigenvalues must be added with $k_0^2+k_3^2$ to obtain the spectrum, 
as done in eq. (\ref{hom}).

\subsection{Straight thick center vortices}
\label{straight}

We will first review the situation for procedure I. Proposing eigenfunctions of the form 
\begin{equation}
 \psi^{(\rm I)}_m =Z_m(\rho)\, e^{im\varphi},
\end{equation}
we are led to,
\begin{equation}
\left[ -\frac{1}{\rho} \partial_\rho\, \rho\, \partial_\rho + \frac{(m +
y^{(\rm I)}(\rho))^2}{\rho^2}\pm \Upsilon^{(\rm I)}\right] Z_m= -\kappa^2\, Z_m,
\label{eigen}
\end{equation}
\begin{equation}
 y^{(\rm I)}(\rho)=f(\rho)
\makebox[.5in]{,} 
\Upsilon^{(\rm I)}=\frac{2}{\rho} \, \partial_\rho f. 
\end{equation}
As is well-known, vortices turn out to be unstable in this case \cite{cho5,bordag}.
This comes about as the Schr\"odinger problem with
$g_m^{(b)}=2$
in eq. (\ref{eigen}), can be associated with a spin one-half particle with
gyromagnetic ratio four. Then, for the vortex profile ${\cal B}^3_\mu =f(\rho)\,
(1/g)\partial_\mu \varphi$, 
the operator $K^{(\rm I)}_-$, having an attractive potential, contains bound
states with the corresponding negative eigenvalues
\cite{bordagQM,morozQM}. In an example where $f(\rho)=\Theta(\rho-\rho_v)$, i.e.,
\begin{equation}
 y^{(\rm I)}(\rho)=\Theta(\rho-\rho_v)
\makebox[.5in]{,}  \Upsilon^{(\rm I)}=\frac{2}{\rho}
\, \delta(\rho -\rho_v),
\end{equation}
the solution is given by
\begin{equation}
\psi^{(\rm I)}_m =\left\{ \begin{array}{ll}
I_{|m|}(\kappa \rho)\, e^{im\varphi},  & \rho< \rho_v \\
K_{|m+1|}(\kappa \rho)\, e^{im\varphi},& \rho > \rho_v .
\end{array}\right. 
\end{equation} 
Multiplying eq. (\ref{eigen}) by $\rho$, and integrating around $\rho=\rho_v$,
the following constraint is obtained,
\begin{equation}
 x \left[
\frac{I_{|m|+1}}{I_{|m|}}+\frac{K_{p+1}}{K_{p}} \right]_{x=\kappa
\rho_v}=\mp 2 + q ,
\end{equation}
\begin{equation}
 p=|m+1|
\makebox[.5in]{,} q=p-|m|.
\end{equation} 
In the attractive case, this equation has two solutions: $m=0$ (lower
eigenvalue) and $m=-1$. Of course, for the repulsive case ($K_+^{(\rm I)}$) there are no
solutions. 

In procedure II, the problem combines a $g^{(b)}_m=2$ factor for the  ${\cal
B}^3_\mu$-sector and a $g^{(d)}_m=1$ factor for the $C^{(\hat{n})3}_\mu$-sector
representing the frame defect. Then, the spectrum has to be
reexamined. The operator $K^{(\rm II)}_-$ ($K^{(\rm II)}_+$) also contains a singular repulsive (attractive) 
core originated from the defect, besides the extended attractive (repulsive) contribution. That is, using eq. (\ref{SchC}), the eigenvalue problem 
is obtained from that given in eq. (\ref{eigen}), replacing $y^{(\rm I)}(\rho)$, $\Upsilon^{(\rm I)}$  by 
\begin{equation}
 y^{(\rm II)}(\rho)=f(\rho)-1
 \makebox[.5in]{,}
\Upsilon^{(\rm II)}=\frac{1}{\rho}
[2\, \partial_\rho f -\delta(\rho)].
\label{nreg}
\end{equation}
The singular part can be treated by means of a regularization accompanied by 
a proper renormalization procedure, if necessary. The latter requirement is needed in problems where a thin vortex is coupled with 
gyromagnetic factor $g_m >1$. When the potential part is attractive, the delta produces bound states with (negative) divergent eigenvalues. This can be 
overcomed by also tending $g_m\to 1$ so as to stabilize the bound states at a finite value \cite{bordagQM}. 
An alternative way, is based on the many different manners to define the singular problem such that the associated operator be self-adjoint. 
The possible extensions are physically inequivalent and can be labeled in terms of the possible behaviors of the wavefunction at the origin.  
The appropriate self-adjoint extension to be used must be fixed on physical grounds. However, this method is not always equivalent to 
the regularization/renormalization process \cite{bordagQM}. In our case, the problem will be defined by means of a regularization, and as 
the frame defect originates a contribution coupled with $g_m=1$, the critical gyromagnetic ratio,
no renormalization will be needed in this case. Let us consider again $f(\rho)=\Theta(\rho-\rho_v)$ and a regularized version of (\ref{nreg}),
\begin{equation}
 y^{(\rm II)}(\rho)=\Theta(\rho-\rho_v)-\Theta(\rho-a)
 \makebox[.3in]{,}
\Upsilon^{(\rm
II)}=\frac{1}{\rho} [2\, \delta(\rho -\rho_v) -\delta(\rho-a)] 
\label{reg}
\end{equation}
(with $a \to 0$). In the case of $K^{(\rm II)}_-$, we can propose a general eigenfunction of the form
\begin{equation}
 \psi^{(\rm II)}_m =X_m\, e^{im\varphi}.
\end{equation}
Using the constraints imposed by the delta potentials at $\rho=\rho_v$ and $\rho=a$, we have checked that for any $\rho> a$, and after taking the limit 
$a\to 0$, the possible solutions are given by, $X_m = Z_m(\rho)\, e^{i\varphi}$,
\begin{equation}
 \psi^{(\rm II)}_m =\left\{ \begin{array}{ll}
I_{|m|}(\kappa \rho)\, e^{i(m+1)\varphi},  & \rho< \rho_v \\
K_{|m+1|}(\kappa \rho)\, e^{i(m+1)\varphi},& \rho > \rho_v ,
\end{array}\right. 
\end{equation}
with $m=0,-1$. That is, the eigenfunctions are related with those obtained in procedure I,
in the form $\psi^{(\rm II)}_m =\psi^{(\rm I)}_m\, e^{i\varphi}$. However, noting that for $\rho < \rho_v$ these solutions are $I_0(\kappa \rho)\, e^{i\varphi}$
and $I_1(\kappa \rho)$, we see that the $m=0$ solution cannot be accepted as, in the limit $\rho \to 0$, 
$I_0(\kappa \rho)$ is nonvanishing and $e^{i\varphi}$ is ill-defined. This is in contrast to what happens in procedure I where both 
behaviors $I_0(\kappa \rho)$, $I_1(\kappa \rho)e^{-i\varphi}$ are acceptable. Therefore, in this example we see that the presence of the frame defect 
reduces the number of bound states in one. This is also expected to occur whenever $f(\rho)$ is well suppressed in a region $\rho < \rho_v$. 

With regard to $K_+^{(\rm II)}$, we have an Aharonov-Bohm type sector with an attractive
singular potential coupled with $g^{(d)}_m=1$, and a repulsive extended part 
coupled with $g^{(b)}_m=2$. If we disregard the
extended part and regularize the singular part with a finite size, as is well known, the corresponding spectrum is formed by a
continuum of eigenvalues extending from $0$ to $+\infty$, plus
Aharonov-Casher localized zero modes that occur for fluxes $2\pi\alpha /g$, with
$\alpha \geq 2$ \cite{a-c}. Then, for an elementary center vortex ($\alpha=1$),
neither bound states nor zero modes are expected. 
In the above example, this has been verified by considering the regularization (\ref{reg}),
and checking that there are no regular solutions to the constraints imposed by
the delta potentials at $\rho=\rho_v$ and $\rho=a$.

In the previous discussion, two important points that deserve a more careful analysis arise. 
First, regarding the boundary conditions at the
defect. Why the ill-defined mode of $K_-^{(\rm II)}$ should be precluded? After all, similar
configurations are generally accepted in connection with Abelian Dirac strings.
The answer precisely relies on the non Abelian nature of the underlying fields.
The class of fields to be considered in the integration measure is closely
related to the class of singular gauge transformations that can be implemented
in the initial theory. For instance, in the case of 
$SU(2)$, phase factors $e^{\pm i 2\varphi}$ are naturally generated in the
charged sector when considering topologically trivial singular gauge transformations 
$U$, introducing closed Dirac strings. These transformations can be
continuously deformed to the identity map, so that these factors should be
considered as harmless and naturally acceptable in the integration measure for
the charged fields. On the other hand, due to the ideal center
vortex, there is no (singular) gauge transformation that can generate a phase
factor $e^{\pm i\varphi}$. This point will be carefully analyzed in section \S 
\ref{DC}. 

The second point is concerning the number of bound states. In ref.
\cite{exca}, under quite general conditions, it has been shown that in a problem with gyromagnetic factor $g_m=2$
(in general, above a critical value that in our notation is $g_m>1$) the number of bound states of $K_-^{(\rm I)}$ is at least $F+1$, where $F\geq 0$ is the integer part
of the total magnetic flux, when written in units of the 
elementary flux. Of course, when $F=0$, the situation for $K_+^{(\rm I)}$ is symmetric and it also has a bound state, while for $F\geq 1$, $K_+^{(\rm I)}$ has no bound states. 
An elementary center vortex is associated with ${\cal B}^3_\mu$ containing one flux quantum, then
the number of bound states in procedure I is at least two (in the example this bound is saturated).  
As a consequence, in procedure II, the number of bound states for a straight
center vortex, taking into account the effect of the $C^{(n)}_\mu$-sector, is at least
one. Note also that the eliminated mode does not present a ``centrifugal'' barrier at small 
$\rho$, so that it is generally expected to be the one corresponding to the lower
eigenvalue (more bounded state); this has been numerically verified in the example. 
Then, in procedure II, a straight line center vortex is still unstable, but in a
milder form. 

\subsection{Thick center vortices formed by a pair of straight components}

Motivated by the line of reasoning given in 
ref. \cite{cho5}, in connection with a pair of axially symmetric
monopole-antimonopole strings, the discussion in the previous subsection opens
the possibility of obtaining a stable configuration by considering, instead of a
vortex localized on a unique straight line, 
a pair with centers at ${\rm F}_-$, ${\rm F}_+$ separated by a distance $2a$,
having magnetic flux with opposite orientations.
In fact, this configuration is more closely related to a center vortex
loop than the previously discussed single straight line component.  
Then, let us first consider procedure I, with a
profile,
\begin{equation}
 {\cal B}^3_\mu= \frac{1}{g} f(\tau)\, \partial_\mu \sigma,
\end{equation}
where $\tau$, $\sigma$ are bipolar coordinates defined by 
\begin{equation}
x_1=a\,\frac{\sinh
\tau}{\cosh \tau - \cos \sigma}
 \makebox[.5in]{,}
x_2=a\,\frac{\sin \sigma}{\cosh \tau - \cos
\sigma}.
\end{equation}
The center ${\rm F}_+$ (${\rm F}_-$) corresponds to $\tau \to +\infty$
($-\infty$) and the multivalued angle $\sigma$ changes by 
$+2\pi$ ($-2\pi$) when we go around it anticlockwise. The profile is suppressed
for $|\tau|> \tau_v$ while it attains the value 
$1$ outside this region. Here, the stability analysis depends on the spectrum of
$K_{\pm}^{(\rm I)}$ in 
eq. (\ref{gm4}), which is similar for both operators. For definitness, we will
consider the $K_{-}^{(I)}$ case, whose eigenvalue problem is, 
\begin{equation}
 -(1/h^2)[\partial^2_\tau + (\partial_\sigma+igf(\tau))^2 + 2 \partial_\tau f]\,
\psi = -\kappa^2 \psi,
\label{bip}
\end{equation}
where $h=a[\cosh \tau -\cos \sigma]^{-1}$ is the scale factor for both
coordinates. For the discrete spectrum we have to ask for a finite $\int d\tau
d\sigma\, h^2 \bar{\psi}\psi$. Equation (\ref{bip}) is not separable, however, we
can propose the general expansion
\begin{equation}
 \psi = \sum_m Z_m (\tau)\, e^{im\sigma},
\end{equation}
and
note that when we approach the center vortices (when $|\tau|$ increases), the scale factor rapidly becomes $h \approx a \,
e^{-\tau}$ ($a \, e^{\tau}$), for positive (negative) $\tau$. 
In this regime, we can define $\rho_{+}=2a\, e^{-\tau}$, $\rho_{-}=2a\,e^{+\tau}$ to show that close to the centers $F_\pm$, $\psi$ 
satisfies,
\begin{equation}
 \left[-\frac{1}{\rho_\pm} \partial_{\rho_\pm} \rho_\pm \partial_{\rho_\pm} -
\frac{\partial_\sigma^2}{\rho^2_\pm} \right]\, \psi
\approx -\kappa^2 \psi .
\end{equation}
Then, the equation becomes separable, $\psi$ can be approximated by a combination
of Bessel functions, and near the center ${\rm F}_+$ (${\rm F}_-$), where the potential in eq.
(\ref{bip}) 
is attractive (repulsive), we have the behaviors 
\begin{equation}
\psi \approx \left\{ \begin{array}{ll}
\sum_m a_m I_m (2\kappa a\, e^{-\tau})\,
e^{im\sigma} ,  & {\rm near~}{\rm F}_+ \\
\sum_m b_m I_m (2\kappa a\, e^{+\tau})\, e^{im\sigma} ,& {\rm near~}{\rm F}_- .
\end{array}\right. 
\end{equation} 
Note that the eigenvalue problem in eq. (\ref{bip}) has no special symmetry. For example, if
$f(\tau)$ is even, the second term is even, while the third term is odd. So that the coefficients $a_n$
and $b_n$ will be generally different.
It is also important to underline that the result of ref. \cite{exca}, concerning
the number of bound states in magnetic vortex backgrounds, is general, not restricted
to axially symmetric situations. Therefore, in procedure I, the proposed background ($F=0$)
is still unstable (it implies at least one bound state). 

Now, let us follow procedure II, by taking the diagonal deformation (see eq. (\ref{thickvII})) of a thin center vortex defined
by the local frame,
\begin{equation}
 \hat{n}_a= e^{i\sigma M^3}\, \hat{e}_a
 \makebox[.5in]{,}
\hat{n}_3= \hat{e}_3. 
\end{equation}
The angle $\sigma$ is only multivalued at ${\rm F}_\pm$ ($\tau \to \pm \infty$). For
instance, when $\tau$ tends to infinity, $\sigma$ approaches the polar angle with respect to ${\rm F}_+$. Then, for any finite $\tau$
the frame defect is only manifested in the kinetic part of $K_{-}^{(\rm II)}$ in eq. (\ref{SchC}), as a simple calculation leads to
\begin{equation}
 C_\mu^{(\hat{n})3}=-\frac{1}{g}\hat{n}_1\cdot  \partial_\mu \hat{n}_2 =
-\frac{1}{g}\partial_\mu \sigma,
\end{equation}
that implies $\tilde{H}^3_{12}(C^{(\hat{n})})=0$ whenever $\sigma$
is single-valued.
That is, for finite $\tau$, the only effect of the local frame is producing the shift 
$f(\rho) \rightarrow f(\rho)-1$ in the second term of eq. (\ref{bip}). Then, if
a bound state exists, it should be of the form $\psi\,e^{i\sigma}$, where $\psi$ is the bound 
state of procedure I, that will be unique for an appropriate profile 
$f(\tau)$ and distance $2a$. In this case, $\psi$ will necessarily be the fundamental state, that must
minimize the expectation value of $K_-^{(\rm I)}$. As a consequence, the coefficient $a_0$ is expected to be nonzero, as it is associated 
with the mode $I_0(\kappa \rho_+)$ that does not see the centrifugal barrier at $\rho_+ \to 0$ ($I_0(0)\neq 0$). 

In other words, the bound state $\psi$ of $K_-^{(\rm I)}$ will tend to a nonzero value ($\psi \to a_0 \neq 0$)  
when the attractive center is approached (this behavior is compatible with square integrability as $h$ is exponentially 
suppressed in that limit). For this reason, the mode $\psi \, e^{i\sigma}$, the only wave function 
that could be a bound state for $K_-^{(\rm II)}$, cannot be accepted; it contains an ill-defined
phase factor at the center ${\rm F}_+$. Summarizing, the frame defect present in our thick center vortex ansatz stabilizes the whole configuration.
With regard to scattering, those solutions that in procedure I vanish at ${\rm F}_\pm$ will be well behaved solutions
of the operators $K^{(\rm II)}_\pm$ (after including the factor $e^{i\sigma}$), forming its continuum spectrum.
This amounts to a correlation between the charged fields and the location of the frame defects, where they must vanish.

\section{Frame defects, boundary conditions and correlations}
\label{DC}

In order to have a better understanding of the relationship between the acceptable singularities 
for the charged fields and the boundary conditions for the eigenvalue problem
characterizing fluctuations, let us consider thin configurations. 
For instance, using the $SU(2)$ mapping  $S(\varphi)=e^{i\varphi \, T^3}$ in eq. (\ref{examp-simp}), we get the thin object,  
\begin{equation}
 \vec{A}^{\,\rm{thin}}_\mu \cdot \vec{T}
=\left[\frac{1}{g}\,
\partial_\mu \varphi +{\cal A}^3_\mu \right]\, \hat{e}_3 + {\cal A}^1_\mu\,
\hat{n}_1 + {\cal A}^2_\mu\, \hat{n}_2 
\makebox[.5in]{,}
\hat{n}_a= e^{i\varphi M^3} \hat{e}_a .
\label{su2examp}
\end{equation}
Then, we see that the fluctuation operator in this case can be obtained from the one computed in eq. (\ref{SchC}),
for the diagonally deformed ansatz in eq. (\ref{thickvII}), by simply replacing ${\cal B}^3_\mu \rightarrow \frac{1}{g}\,
\partial_\mu \varphi $, thus giving,
\begin{equation}
K_\pm = - \nabla^2 \pm \Upsilon
\makebox[.5in]{,}
\Upsilon=\frac{2\pi}{g} \delta^{(2)}(x,y).
\end{equation}
These operators only contain a delta potential and in the regularized problem, by means of a
tube with localization radius $\epsilon$,
$K_-$ has a discrete spectrum that is found to scale as $-1/\epsilon^2$, due to scale invariance of the initial
problem \cite{b-f}. This spectrum would lead to
instability, but towards what kind of configuration? 
For example, in procedure I, where center vortices are introduced in terms of a
background field ${\cal B}_\mu^r$ and a trivial globally defined color frame
$\hat{e}_A$, instability is towards a trivial background obtained by
expanding the vortex radius $\rho_v$ to infinity. 
To have an idea about the possibilities for the thin center vortex, it is
instructive to analyze the case where in the previous example, $S(\varphi)=e^{i\varphi \, T^3}$ is replaced by
$U(\varphi)=e^{2i\varphi \, T^3}$, to obtain,
\begin{equation}
\left[\frac{2}{g}\, \partial_\mu \varphi +{\cal
A}^3_\mu \right]\, \hat{e}_3 + {\cal A}^1_\mu\,
\hat{n}_1 + {\cal A}^2_\mu\, \hat{n}_2 
\makebox[.5in]{,}
\hat{n}_a= e^{2i\varphi M^3} \hat{e}_a. 
\label{Diracon}
\end{equation}
The analysis of quadratic fluctuations would be
similar to the previous one, with the replacement $\Upsilon \rightarrow \frac{4\pi}{g}
\delta^{(2)}(x,y)$, and the obtained configuration seems to be unstable.
However, unlike $S$ that is discontinuous on the three volume $\vartheta$ defined by the positive $x_1$-axis, 
for every $x_0$, $x_3$ (see the discussion in \S \ref{diagdef}), $U$ has no discontinuity ($U(0)= U(2\pi)= I$), so that no ideal center vortex part would be present in the 
representation given in (\ref{usual-thin}) (with $S\rightarrow U$). Therefore, the configuration in (\ref{Diracon}) can be also represented as,
\begin{equation}
U {\cal A}^A_\mu\, T^A U^{-1}+\frac{i}{g} U\partial_\mu U^{-1}.
\end{equation}
Of course, this corresponds to the introduction of a closed Dirac
string by means of a topologically trivial singular gauge transformation $U$,
that can be continuously deformed (together with the associated local frame) to
a trivial configuration $U \rightarrow I$; note that the first homotopy group of
$SU(2)$ is trivial and, when we go around the thin center vortex, $U$ defines a closed loop as it is single-valued. The above mentioned
deformation is not along the diagonal direction, it uses the whole non Abelian
nature of the group. In the path integral, if we analyze the situation before
the quadratic fluctuations were considered, the closed Dirac string can be eliminated by 
means of an appropriate change of variables. In terms of the fields, this amounts to a $(2/g)\,
\partial_\mu \varphi$  shift of the diagonal sector accompanied by an
$e^{-2i\varphi}$ phase transformation of the charged sector. Then, the
instability seemingly implied when analyzing quadratic fluctuations is an
artifact of the approximation. 

This is in contrast to what happens in the case of thin center vortices. Because
of the ideal part in eq. (\ref{usual-thin}), the thin center vortex in eq. (\ref{su2examp}) has nothing
to do with a (gauge) transformation of the fields. 
Given a thin center vortex location, there is no manner to deform the local frame
so as to produce a trivial configuration. For $N=3$ (resp. $N=2$), when we go
around the center vortex worldsheet, the pair of basis elements $(\hat{n}_4$,
$\hat{n}_5)$, $(\hat{n}_6$, $\hat{n}_7)$ (resp. $\hat{n}_1$, $\hat{n}_2$) rotate
once, leaving $\hat{n}_A$, $A=1,2,3,8$ (resp. $\hat{n}_3$) invariant. In other
words, in this case there is an obstruction to deform the configuration, possed
by the defects of the local color frame, and there is no field transformation on
the initial variables to absorb the defect. That is, in this case we have a
genuine instability problem. 
A possible way out is to start with a new configuration where the diagonal
sector is deformed to contain a thick profile, and the local frame is maintained
in the charged sector, thus defining the type II thick center vortices proposed
in this article. 

Summarizing, in Yang-Mills theories, the topologically trivial $U$-con\-fig\-u\-ra\-tions are to be considered as trivial gauge transformations
introducing closed Dirac worldsheets. Then, the space of charged fields must be closed under
$e^{\pm 2i\varphi}$ phase transformations, and no correlation must exist between
the frame defect (Dirac worldsheet) and the charged fields. It is clear that this
should not be the case for the type II thick center vortices. They have nothing
to do with gauge transformations, so that $e^{\pm i\varphi}$ phases in the charged sector should not be considered as an artifact,  
and ill-defined expressions at the origin should be avoided by requiring a correlation: 
charged fields should be zero at the frame defect in type II center vortices. The presence or absence of correlations is at the basis
of the observability/nonobservability of the associated objects, that is, the possibility of making changes of
variables so as to decouple them or not in the partition function
(see ref. \cite{aom}). For a similar discussion involving correlations between
Wilson surfaces and charged dual fields, see ref. \cite{Lucho1}.

\section*{Conclusions}
\label{conc}

In scenarios based on center vortex degrees of freedom, a part of the problem of
confinement amounts to show how can center vortices be relevant stable objets in the infrared regime, characterized by physical properties such as
their thickness and stiffness. As is well-known, when these objects are defined as usual, only in terms of a background field, they are unstable. 
This is due to the gyromagnetic ratio $g^{(b)}_m=2$ that couples off-diagonal gluons with background fields. The value for this ratio is related with
general high energy properties of scattering \cite{weinberg} and plays a crucial role in perturbative Yang-Mills theories. It 
leads to a paramagnetic vacuum that due to Lorentz covariance is related with asymptotic freedom (antiscreening) \cite{sno,W,polya1}.

In this work, we have initially rewritten the {\it thin} center vortices, usually defined in continuum $SU(N)$ gauge theories, 
only in terms of a local color frame $\hat{n}_A$, thus
implementing in a natural way the subtraction of the ideal vortex part. This comes about as $\hat{n}_A$ transforms in the adjoint 
representation of $SU(N)$, so that it is single-valued along any loop. In the case of $SU(2)$, this procedure has
been discussed in ref. \cite{lucho}, where we have noted that frame defects can be used not only to describe monopoles \cite{cho-a}-\cite{Shaba}, but also center vortices. 
In that reference, by supplementing the coupling between defects and quantum
fluctuations with an effective action for the former, plausible models
containing gluon, monopole and center vortex effective fields were obtained
\cite{lucho}. This also provided a natural framework in the continuum to discuss, in terms of large dual transformations,
phases where the Wilson surface can be decoupled vs. phases where it becomes a dynamical variable, signaling
confinement \cite{Lucho1}.

The thin center vortex representation we obtained here permitted to define a thick object as a diagonal deformation of the former. 
This is in fact quite natural and amounts to simply considering a thick profile in the diagonal sector, 
maintaining the local frame in the charged sector, together with its associated defect. The later ingredient is not present in the usual thick center vortex definition, and is in fact welcomed because of the following reasons.

In both alternatives, when the fields that represent the trivial sector are small (i.e. $q^A_\mu$, ${\cal A}^A_\mu$, 
for procedures I, II, resp.), the Wilson loop is close to a center element. However, in our procedure, the Wilson 
loop for {\it any} trivial configuration ${\cal A}^A_\mu\, \hat{e}_A$ and one that contains a diagonally deformed 
thin center vortex on top of it, always differ by a center element, when the Wilson loop passes outside the associated core.

The local color frame confers to the anstaz some topological features typical of the adjoint representation of $SU(N)$. For instance, in the case 
of $SU(2)$ this corresponds to the $SO(3)$ group defining the local color frame $\hat{n}_A$, whose first homotopy group is $Z(2)$. This topological aspect, 
manifested as a nontrivial frame defect coupled with gyromagnetic ratio $g^{(d)}_m=1$, should give a better behavior regarding stability than in the usual ansatz only based on a background field. 

In refs. \cite{bordag,diakonov-center}, based on a general argument relying on renormalizability, it has been shown that,
up to one loop, the total energy for a magnetic object of size $\rho_0$ is always negative, after 
minimizing with respect to $\rho_0$. For example, the particular form of a classical background is only manifested in the depth at the minimum. 
However, as noted in \cite{bordag}, background field configurations are in fact unstable. In all the examples considered, 
the fluctuation operator has bound states so that the total energy contains an imaginary part. 
This is in agreement with a theorem stating that a Schr\"odinger operator for a spin-$1$ particle coupled to a magnetic background with gyromagnetic ratio $g^{(b)}_m=2$ (in general, above the critical value $g_m>1$) the number of bound states is at least $F+1$, where $F$ is the integer part
of the total magnetic flux, when written in units of the 
elementary flux. When $F\neq 0$, this theorem only applies to $K_-^{(\rm I)}$, the operator whose magnetic moment is parallel to the total magnetic flux
(when on average the potential is attractive). 
Then, for a center vortex background ($F=1$),  $K_-^{(\rm I)}$ has at least two bound states; this is indeed the number observed in \cite{bordag}
(see also \S \ref{straight}).  

The diagonally deformed thin center vortex contains a frame defect, besides the thick background profile. Then, because of the 
natural boundary conditions on the charged fields, the number of bound states of $K_-^{(\rm II)}$, the operator governing 
quantum fluctuations in our proposal,
is reduced by one with respect to the usual procedure. Therefore, for a single straight object, the instablity problem is weaker but still present. 

However, for zero flux ($F=0$) magnetic backgrounds, $K_-^{(\rm I)}$ has at least only one bound state, 
and the above mentioned reduction, operated by the frame defect, can lead to no bound states at all in 
$K_-^{(\rm II)}$ (a similar situation occurs with the other fluctuation operator $K_+^{(\rm II)}$).
Then, the frame defect will stabilize a closed vortex formed by a pair of straight components 
with opposite orientations, for an appropriate distance between them, and vortex profile,  
such that the above mentioned bound is saturated.
This suggests that the thick objects, naturally obtained from the diagonal deformation of the {\it thin} center vortex configurations, 
as proposed in this article, 
could represent relevant degrees of freedom in continuum Yang-Mills theories.

\section*{Acknowledgements}
The Conselho Nacional de Desenvolvimento Cient\'{\i}fico e Tecnol\'{o}gico
(CNPq-Brazil) and the Proppi-UFF are acknowledged for the financial support.


\bibliographystyle{h-physrev4}

\end{document}